\documentclass[runningheads]{llncs}
\usepackage{graphicx}
\def\Plus{\texttt{+}}   
\usepackage{amsfonts}
\usepackage{bbm}
\usepackage[colorlinks=true,linkcolor=blue, citecolor=blue, urlcolor=blue]{hyperref}
\usepackage{hhline}
\usepackage[table]{xcolor}
\newcommand\like[1]{\begin{picture}(1,1)
\ifnum0=#1\put(.5,.35){\circle{0.5}}\else
\ifnum10=#1\put(.5,.35){\circle*{0.5}}\else
\put(.5,.35){\circle{0.5}}\put(.5,.35){\circle*{.0.5}}
\fi\fi\end{picture}}
\usepackage{stmaryrd}
\usepackage{amsmath}
\usepackage[misc]{ifsym}

\DeclareMathOperator*{\argmin}{\arg\!\min}
\DeclareMathSymbol{\shortminus}{\mathbin}{AMSa}{"39}

\begin{document}
\title{ModDrop++: A Dynamic Filter Network with Intra-subject Co-training for Multiple Sclerosis Lesion Segmentation with Missing Modalities}

\titlerunning{ModDrop\Plus\Plus{}}
%
\author{Han Liu\inst{1}\Letter\and
Yubo Fan\inst{1} \and
Hao Li\inst{2} \and
Jiacheng Wang\inst{1} \and
Dewei Hu\inst{2} \and
Can Cui\inst{1} \and
Ho Hin Lee\inst{1} \and
Huahong Zhang\inst{1} \and
Ipek Oguz\inst{1,2}}

\institute{Department of Computer Science, Vanderbilt University\and 
Department of Electrical and Computer Engineering, Vanderbilt University\\
\email{han.liu@vanderbilt.edu}}

%
\authorrunning{H. Liu et al.}
%


%
\maketitle              
\begin{abstract}
Multiple Sclerosis (MS) is a chronic neuroinflammatory disease and multi-modality MRIs are routinely used to monitor MS lesions. Many automatic MS lesion segmentation models have been developed and have reached human-level performance. However, most established methods assume the MRI modalities used during training are also available during testing, which is not guaranteed in  clinical practice. Previously, a training strategy termed Modality Dropout (ModDrop) has been applied to MS lesion segmentation to achieve the state-of-the-art performance with missing modality. In this paper, we present a novel method dubbed ModDrop\Plus\Plus{} to train a unified network adaptive to an arbitrary number of input MRI sequences. ModDrop\Plus\Plus{} upgrades the main idea of ModDrop in two key ways. First, we devise a plug-and-play dynamic head and adopt a filter scaling strategy to improve the expressiveness of the network. Second, we design a co-training strategy to leverage the intra-subject relation between full modality and missing modality. Specifically, the intra-subject co-training strategy aims to guide the dynamic head to generate similar feature representations between the full- and missing-modality data from the same subject. We use two public MS datasets to show the superiority of ModDrop\Plus\Plus{}. Source code and trained models are available at \url{https://github.com/han-liu/ModDropPlusPlus}.

\end{abstract}
\section{Introduction}
Multiple Sclerosis (MS) is a chronic inflammatory neurological disease characterized by focal lesions and diffuse neurodegeneration \cite{kolasinski2012combined}. Magnetic Resonance Imaging (MRI) is crucial for the diagnosis and monitoring of MS lesions \cite{kaunzner2017mri}. To characterize different types of MS lesions, multiple MR imaging modalities such as T1w, T2w, Proton Density (PD), Fluid-attenuated inversion recovery (FLAIR), and T1-contrast enhanced (T1CE) are routinely collected in clinical practice. There have been many automatic MS lesion segmentation algorithms in the medical imaging community and some even achieved human-level performances \cite{carass2017longitudinal,valverde2017improving,zhang2019multiple}. Most established methods have been developed based on the assumption that the imaging modalities used during training are available during deployment. However, in clinical practice, it is not possible to always meet this assumption because (1) the number and categories of MRI sequences acquired at different sites can vary and (2) the acquired MR sequences for a specific patient may be unusable due to poor image quality. The modality mismatch between the training and testing data, i.e., the missing modality problem, significantly limits the use of the existing methods during deployment.

A straightforward way to tackle this problem is to train independent models for each missing-modality condition but this requires a long training time. To overcome this issue, a training strategy termed Modality Dropout (\textit{\textbf{ModDrop}}) \cite{neverova2015moddrop} has been developed and widely used in various fields such as computer vision \cite{de2020input}, dialogue systems \cite{sun2021non} and medical imaging \cite{la2020automated,li20183d,van2018learning}. Particularly, for MS lesion segmentation, Feng \textit{et al.} \cite{feng2018self} adopted ModDrop and achieved the state-of-the-art performance to handle MRIs with missing sequences. Similar to the regular dropout for neurons, in ModDrop, some modalities are randomly dropped-out during training and a unified network is trained to create feature representations that are robust to all missing conditions \cite{cheerla2019deep}. ModDrop is easily applicable to any existing models without modifying the network architectures. However, it may suffer from two limitations: (1) regardless of different missing conditions, ModDrop always forces the network to learn a single set of parameters, which may limit the expressiveness of the network, and (2) ModDrop does not leverage the intra-subject relation between full- and missing-modality data.

Our work is heavily inspired by dynamic filter networks \cite{klein2015dynamic} and co-training strategies. In dynamic filter networks, the filter parameters can be dynamically generated conditioned on either an input image \cite{klein2015dynamic}, latent features \cite{hu2021domain}, or a pre-defined task code \cite{zhang2021dodnet}. In medical image analysis, dynamic filter networks have been successfully used to improve the model accuracy and efficiency in a variety of tasks such as image translation \cite{yang2021unified}, partial label training \cite{zhang2021dodnet}, and domain generalization \cite{hu2021domain}. However, the effectiveness of dynamic filter networks on missing modality issue has not been explored. On the other hand, co-training strategies have been widely used for knowledge distillation to minimize the gap between multi- and uni-modalities. For instance, KD-net \cite{hu2020knowledge} can distill knowledge from multi-modal data via co-training to improve the performance of a uni-modal model. However, KD-net suffers from inefficiency because each uni-modal model has to be distilled from the multi-modal model separately.

We propose a novel training method, dubbed \textit{\textbf{ModDrop}}\textbf{\Plus\Plus}, to address missing modalities  for MS lesion segmentation. Our \underline{novel contributions} are as follows: 
\begin{itemize}
    \item We devise a plug-and-play dynamic head to improve the network expressiveness. The dynamic head can adaptively generate the model parameters conditioned on the missing condition. To the best of our knowledge, this is the first study to apply the dynamic filters for the missing modality problem. 
    \item We introduce a co-training strategy to leverage the intra-subject relation between the full- and missing-modality data. 
    \item Built upon the leading method \cite{zhang2019multiple} of the ISBI challenge  \cite{carass2017longitudinal}, our experiments on two publicly available MS datasets show the superiority of ModDrop\Plus\Plus.
\end{itemize}
\begin{figure}[t]
\includegraphics[width=1\columnwidth]{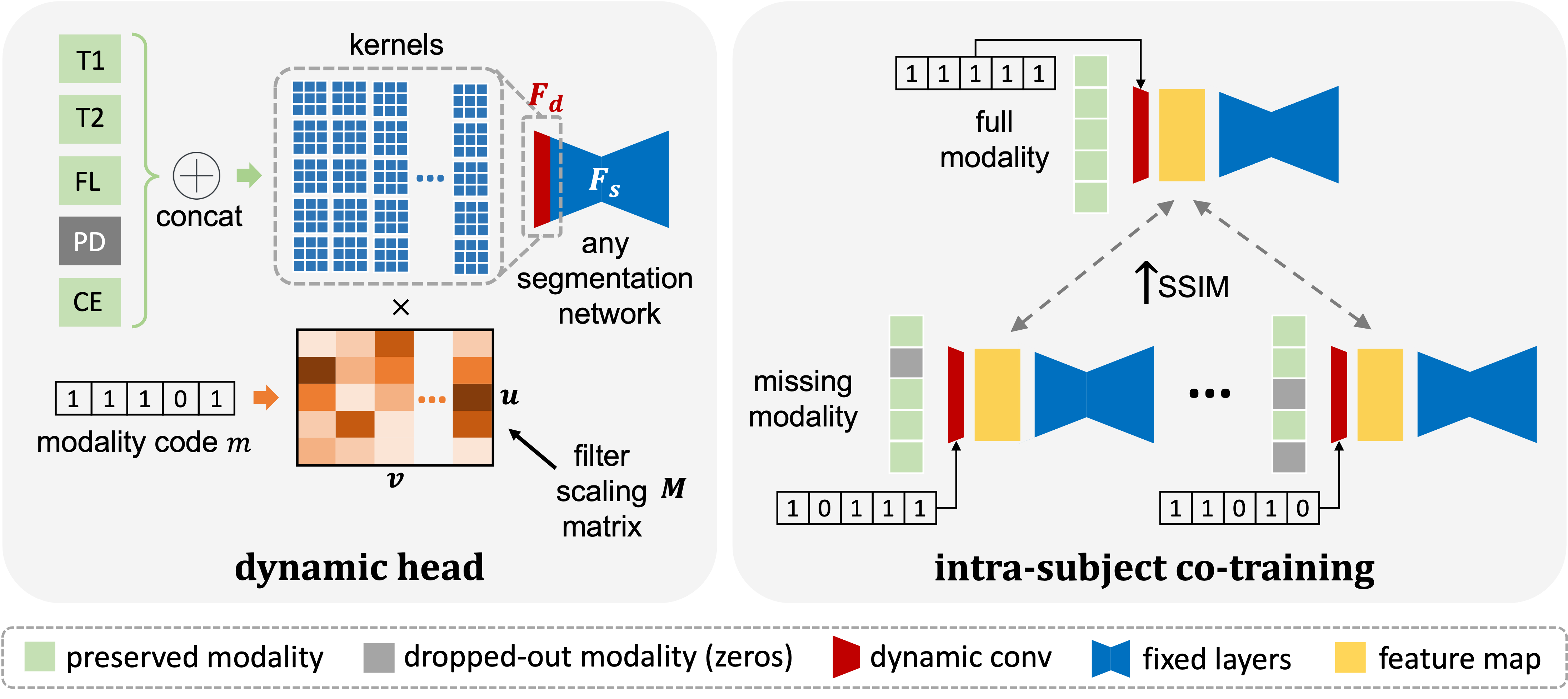}
\centering
\caption{The two key upgrades in ModDrop\Plus\Plus{}: a plug-and-play dynamic head (left) and an intra-subject co-training strategy (right). The dynamic head aims to improve the network expressiveness by learning a set of filter scaling matrices to adaptively adjust the first convolution layer for each missing condition. The intra-subject co-training aims to transfer the knowledge between the full-modality data and the missing-modality data of the same subject, which can guide the dynamic head to produce similar feature representations even when multiple modalities are absent.}
\label{fig1}
\end{figure} 

\section{Methods}
\subsection{Preliminaries}
ModDrop\Plus\Plus{} aims to train a unified MS lesion segmentation model that can be adaptable to an arbitrary number of available MRI sequences (modalities) during deployment. In particular, it inherits the training scheme of the classic ModDrop. Let $x$ and $y$ be the input data of $K$ modalities and the label. During training, a shared network $F$ is trained using the input data with image modalities randomly dropped-out (replaced by zeros). There are $2^K\shortminus1$ unique missing modality configurations, represented by: 


\begin{equation}
    \tilde{x}^{k}=\delta^{k} x^{k},
    k \in \{1,...,K\},
\end{equation}
where 
$\delta^{k}$ is a Bernoulli selector variable that can take on values in $\{0,1\}$ for each modality $k$. The frequency of each possible configuration $i \in \{1,...,2^{K}\shortminus1\}$ being activated during training is determined by the probability $p^{k}$ for each $\delta^{k}$.
Let $\theta_{s}$ be the model parameters of the shared network and $L_{t}$ be the loss function for the downstream task. The learning objective of the traditional ModDrop can be expressed as 
\begin{equation}    \argmin_{\theta_{s}} L_{t}(y,F(\tilde{x}_i|\theta_{s}))
\end{equation}

In this work, we build upon ModDrop and upgrade it to ModDrop\Plus\Plus{} by introducing (1) a plug-and-play dynamic head and (2) an intra-subject co-training strategy, as illustrated in Fig.\ \ref{fig1}. The details of these upgrades are as follows.

\subsection{Dynamic Head with Filter Scaling}
In ModDrop, a single set of model parameters $\theta$ is learned to handle all possible missing conditions. This may limit the expressiveness of the network and lead to suboptimal performance for competing missing conditions. This can get severe with an increasing number of total modalities, since $K$ modalities leads to $2^K\shortminus1$ missing conditions. As shown in Fig.\ \ref{fig1} (left), to improve the network expressiveness, we devise a dynamic head $D$ to adaptively generate model parameters conditioned on the availability of input modalities. We use a binary modality code $m\in\mathbb{R}^K$ where 0/1 represent the absence/presence of each modality. 

To mitigate the large input variation caused by artificially zero-ed channels, we use the dynamic head to generate the parameters for the first convolutional layer $F_{d}$. Suppose the number of input channels and output channels are $u$ and $v$, and the kernel size is $p\times q$. Typically, all parameters in the dynamic convolution layer are generated individually based on a given prior. In our scenario, the dynamic head is asked to learn to generate a total number of $uvpq+b$ parameters from the modality code $m$ ($b$ parameters for bias). However, this mapping might be too difficult to learn for the dynamic head, and our networks fail to converge in preliminary experiments. 

To address this issue, we propose to update our dynamic filters with a filter scaling strategy, which was originally designed for unsupervised image-to-image translation \cite{alharbi2019latent}. Our goal is that kernels should contribute differently for inputs with different dropped-out modalities. The task of our dynamic head then becomes to learn to generate a filter scaling matrix $M\in \mathbb{R}^{u\times{v}}$ for each missing condition, where each element in $M$ represents the contribution (scale) of the corresponding kernel at each missing condition. The kernel weights in $F_{d}$ are updated with the corresponding scale factor by a scalar multiplication. The kernel biases are updated in the same manner. This filter scaling strategy reduces the number of learnable parameters from $uvpq+b$ to $uv+b$, which preserves the dynamic nature of our network but with a much simpler learning task.

\subsection{Intra-subject Co-training}
 With our dynamic head, our network can be considered to have two components: the first convolutional layer $F_{d}$ with dynamic filters and the subsequent fixed layers $F_{s}$ with static filters. The feature map obtained by $F_{d}$ can thus be expressed as $f_{i}=F_{d}(\tilde{x}_{i}|m_{i})$, which are further passed to $F_{s}$ to obtain the final prediction. One assumption of our co-training strategy is that networks can perform the best when fed with full-modality data $x$. Based on this assumption, we can improve the segmentation performance for the missing-modality data $\tilde{x}_i$ when $f_{i}$ becomes similar to $f=F_{d}(x|\mathbbm{1})$, because $f$ and $f_{i}$ are passed to the same network $F_{s}$ with static filters. Motivated by this, we develop an intra-subject co-training strategy to guide the dynamic head to learn to maximize the similarity between $f$ and $f_{i}$. Specifically, we forward the full-modality data $x$ and missing-modality data $\tilde{x}_i$ from the same subject to $F_{d}$ in parallel to obtain the associated feature maps $f$ and $f_{i}$, which are further passed to $F_{s}$ to obtain the final outputs. Besides the downstream task loss $L_{t}$ for both outputs, we also optimize a similarity loss $L_{sim}$ to maximize the similarity between $f$ and $f_{i}$ (Fig.\ \ref{fig1}, right panel). Therefore, the overall training objective of ModDrop\Plus\Plus{} can be expressed as follows:
 
 \begin{equation}
    \argmin_{\theta_{s}, \theta_{d}} 
    \alpha{}L_{t}(y, F_{s}(f))
    +\beta{}L_{t}(y, F_{s}(f_{i}))
    -\gamma{}L_{sim}(f,f_{i})
\end{equation}
where $\alpha$, $\beta$ and $\gamma$ are the coefficients to weigh the importance of each loss term.

Specifically, ModDrop\Plus\Plus{} uses the Structural Similarity Index (SSIM) loss as $L_{sim}$. The SSIM has been widely used as a quality metric for the similarity between two images \cite{hore2010image}. Recently, Wang \textit{et al.} \cite{ssimloss} showed that the SSIM can also be an effective loss to measure the similarity in feature space. The features that we aim to maximize the similarity of are from $F_{d}$ and thus contain mostly low-level structural features, which can be nicely captured by the SSIM loss. 

\subsection{Implementation Details}
Without loss of generality, we develop ModDrop\Plus\Plus{} based on the public implementation of the method leading the ISBI challenge leaderboard \cite{carass2017longitudinal}. This model uses a 2D fully convolutional densely connected network with concatenated 2.5D stacked slices from each modality as the input \cite{zhang2019multiple}. Our dynamic head is a 2D convolutional layer with kernel size of $1\times{1}$. During training, we use the focal loss \cite{lin2017focal} for lesion segmentation, which has been shown to achieve higher Dice coefficient than other losses \cite{zhang2019multiple}. The focusing parameter and the weighting factor of class imbalance are set as 0.25 and 2 respectively as found in \cite{zhang2019multiple}. For intra-subject co-training, we use the SSIM loss with a window size of $11\times{11}$ as the similarity loss. 

In our overall loss function, the weighting factors $\alpha$, $\beta$ and $\gamma$ are empirically set as 1, 1 and 0.05. As in \cite{zhang2019multiple}, we set the initial learning rate as 0.25 (linearly decayed after 100 epochs) with a batch size of 16. All the networks are totally optimized to 300 epochs with an Adam optimizer with a momentum of 0.5.



\begin{figure}[t]
\includegraphics[width=1\columnwidth]{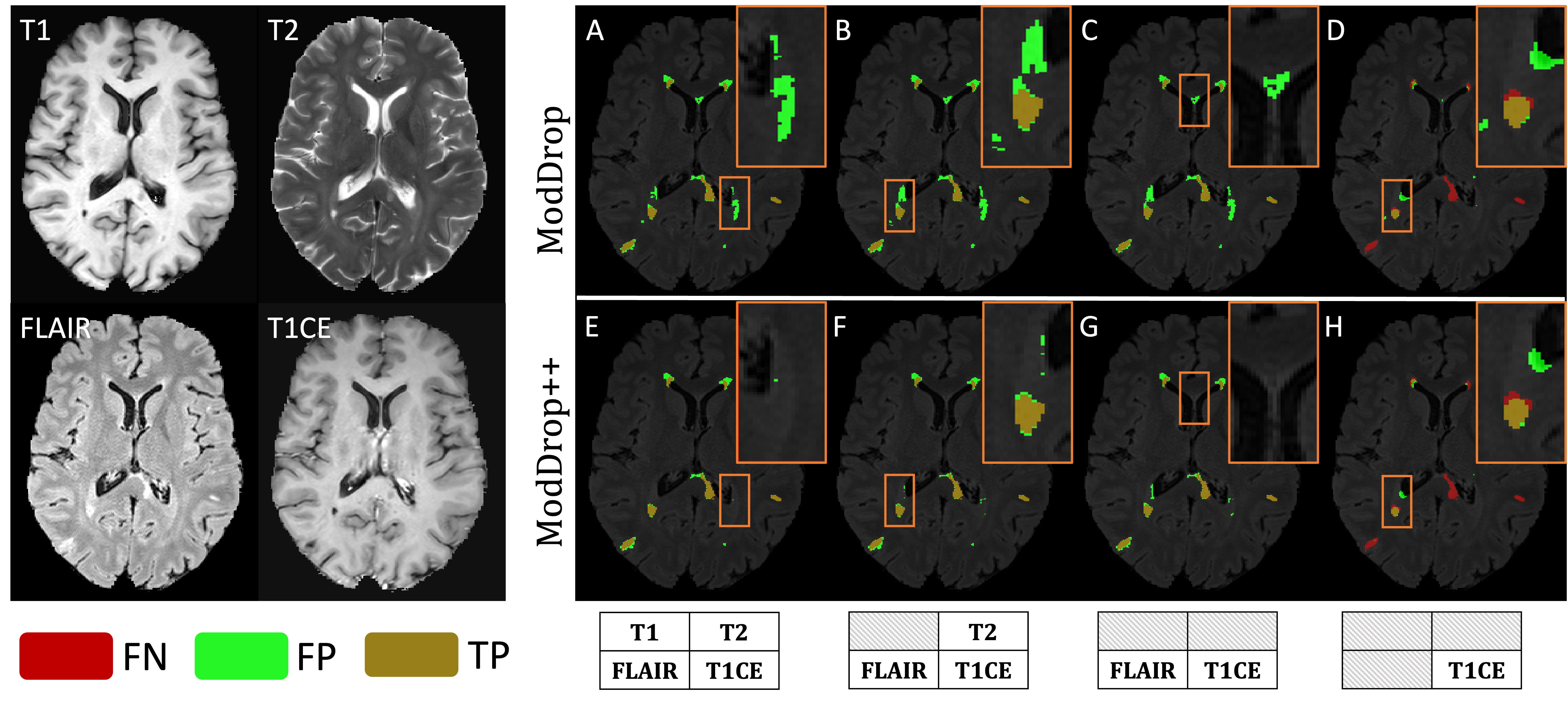}
\centering
\caption{Qualitative comparisons of ModDrop and ModDrop\Plus\Plus{} (ours). The left panel displays the four available modalities of a testing sample from the UMCL dataset. The right panel shows the segmentation results by ModDrop (row 1) and ModDrop\Plus\Plus{} (row 2). The major differences between ModDrop and ModDrop\Plus\Plus{} are zoomed in on the upper-right panels. The number of available modalities from left to right column decreases from 4 to 1.}
\label{fig2}
\end{figure}

\subsection{Datasets and Evaluation Metrics} 

In our experiments, two public 3T MS datasets are used. \textbf{UMCL Multi-rater Consensus Dataset \cite{lesjak2018novel}.} This dataset, referred to as \textbf{UMCL} (University Medical Center Ljubljana), is a cohort of 30 patients with 3D FLAIR, 2D T1w, 2D T1CE and 2D T2w scans. The manual segmentation is delineated by three experts and a consensus was reached via discussion. \textbf{Longitudinal MS Lesion Segmentation Challenge (ISBI) \cite{carass2017longitudinal}.} This dataset contains 21 timepoints from 5 subjects with T1w, T2w, FLAIR, and PD, and two separate manual delineations. For both datasets, we follow the same pre-processing procedures as in \cite{zhang2019multiple} by performing intensity normalization for each image with kernel density estimation. We randomly split the dataset subject-wise with a ratio of 4:1 for training and testing.

We use the evaluation metrics in the ISBI challenge \cite{carass2017longitudinal} including Dice similarity coefficient (DSC), positive predictive value (PPV), true positive rate (TPR), lesion-wise false positive rate (LFPR), lesion-wise true positive rate (LTPR), volume difference (VD), and volume correlation (Corr). To evaluate the overall performance, we follow \cite{carass2017longitudinal} and compute the overall score (SC) as $
    SC=\frac{DSC}{8}+\frac{PPV}{8}+\frac{1-LFPR}{4}+\frac{LTPR}{4}+\frac{Corr}{4} $.

\begin{table}[t]
\centering
\caption{UMCL results. Dice and overall score (SC) are presented for all possible configurations of MRI modalities being either absent (\like{0}) or present (\like{10}).  ModDrop (MD), ModDrop with only the dynamic head (MD\Plus{}) and ModDrop\Plus\Plus{} (MD\Plus\Plus{}) are unified models, while independent models (IM) are trained specifically for each missing condition. The best and second best performances are denoted by \textbf{bold} and \underline{underline}.}
\rowcolors{2}{gray!25}{white}
\begin{tabular}{|cccc|cccc|cccc|}
\hline
\multicolumn{4}{|c}{Modalities} & \multicolumn{4}{|c}{DSC($\uparrow$)} & \multicolumn{4}{|c|}{SC($\uparrow$)} \\ \hhline{|====|====|====|}
\multicolumn{1}{|c}{T1} & \multicolumn{1}{c}{Flair} & \multicolumn{1}{c}{T2} & \multicolumn{1}{c|}{CE} & \multicolumn{1}{c}{MD} & \multicolumn{1}{c}{MD\Plus{}} & \multicolumn{1}{c}{MD\Plus\Plus{}} & \multicolumn{1}{c|}{IM} & \multicolumn{1}{c}{MD} & \multicolumn{1}{c}{MD\Plus{}} & \multicolumn{1}{c}{MD\Plus\Plus{}} & \multicolumn{1}{c|}{IM} \\
\hline
\like{10} & \like{10} & \like{10} & \like{10} & 0.633 & 0.663 &  \textbf{0.704} & \underline{0.690} & 0.610 & 0.633 &  \textbf{0.676} & \underline{0.652} \\
\like{0}  & \like{10} & \like{10} & \like{10} & 0.611 & 0.652 &  \textbf{0.706} & \underline{0.698} & 0.596 & 0.624 &  \underline{0.673} & \textbf{0.674} \\
\like{10} & \like{0}  & \like{10} & \like{10} & 0.400 & 0.453 &  \underline{0.474} & \textbf{0.512} & 0.401 &  \underline{0.443} & 0.442 & \textbf{0.472} \\
\like{10} & \like{10} & \like{0}  & \like{10} & 0.641 & \underline{0.659} &  \textbf{0.693} & 0.649 & 0.623 & 0.629 &  \textbf{0.665} & \underline{0.663} \\
\like{10} & \like{10} & \like{10} & \like{0}  & 0.610 & 0.639 &  \textbf{0.705} & \underline{0.697} & 0.599 & 0.625 &  \textbf{0.674} & \underline{0.661} \\
\like{0}  & \like{0}  & \like{10} & \like{10} & 0.377 & 0.439 &  \underline{0.458} & \textbf{0.482} & 0.379 & 0.430 &  \underline{0.431} & \textbf{0.462} \\
\like{0}  & \like{10} & \like{0}  & \like{10} & 0.629 & 0.653 &  \textbf{0.696} & \underline{0.692} & 0.600 & 0.621 &  \underline{0.671} & \textbf{0.676} \\
\like{0}  & \like{10} & \like{10} & \like{0}  & 0.613 & 0.651 &  \underline{0.704} & \textbf{0.711} & 0.589 & 0.618 &  \textbf{0.676} & \underline{0.655} \\
\like{10} & \like{0}  & \like{0}  & \like{10} & 0.325 & 0.329 &  \underline{0.331} & \textbf{0.382} & 0.303 & 0.338 &  \underline{0.343} & \textbf{0.400} \\
\like{10} & \like{0}  & \like{10} & \like{0}  & 0.382 & 0.427 &  \underline{0.451} & \textbf{0.471} & 0.399 & 0.424 &  \underline{0.427} & \textbf{0.459} \\
\like{10} & \like{10} & \like{0}  & \like{0}  & 0.631 & 0.650 &  \textbf{0.696} & \underline{0.683} & 0.600 & 0.622 &  \textbf{0.670} & \underline{0.658} \\
\like{10} & \like{0}  & \like{0}  & \like{0}  & 0.275 & 0.304 &  \underline{0.324} & \textbf{0.400} & 0.288 &  \underline{0.326} & 0.318 & \textbf{0.398} \\
\like{0}  & \like{10} & \like{0}  & \like{0}  & 0.612 & 0.642 &  \underline{0.693} & \textbf{0.702} & 0.576 & 0.607 &  \underline{0.665} & \textbf{0.670} \\
\like{0}  & \like{0}  & \like{10} & \like{0}  & 0.338 & 0.412 &  \underline{0.420} & \textbf{0.459} & 0.341 &  \underline{0.411} & 0.403 & \textbf{0.456} \\
\like{0}  & \like{0}  & \like{0}  & \like{10} & 0.270 &  \underline{0.304} &  \underline{0.304} & \textbf{0.362} & 0.292 &  \underline{0.333} & 0.330 & \textbf{0.367} \\
\hline
\end{tabular}
\end{table}

\section{Experimental Results}
We compare ModDrop\Plus\Plus{} and ModDrop, which has been demonstrated to be very effective for our application \cite{feng2018self}, in all possible configurations of available MRI sequences. We assess the impact of the dynamic head and the co-training strategy by evaluating the performance of ModDrop with dynamic head but without intra-subject co-training. We also compare the performances of the unified models against independent models trained for each specific missing condition. The independent models are trained with consistent input modalities and are thus more likely to outperform the unified models. These models can be considered as the upper bounds for each missing condition.

\textbf{Quantitative Results.} In Tables 1 and 2, we show the DSC and SC of the compared methods for all possible combinations of MRI modalities, on the UMCL and ISBI 2015 datasets respectively. Remaining evaluation metrics can be found in the supplemental materials. First, we observe that the lesion segmentation performance for all compared methods drops considerably when FLAIR is absent. This is expected because FLAIR is indeed important for lesion identification tasks in real clinical settings and the most commonly used modality for manual lesion delineation. On both datasets, by adding the dynamic head alone (ModDrop\Plus{}), the performances of all missing conditions are improved \textit{consistently} (except when only T1w is available on the ISBI dataset). Moreover, when combining the intra-subject co-training with the dynamic head (ModDrop\Plus\Plus{}), the performance can be further improved, demonstrating the effectiveness of the co-training strategy. Lastly, we observe that the performance gap between the unified model and independent models are substantially minimized by upgrading ModDrop to ModDrop\Plus\Plus{}. 


\textbf{Qualitative Results.} In Fig.\ \ref{fig2}, we qualitatively compare ModDrop\Plus\Plus{} (bottom row) against ModDrop (top row) with different number of available modalities. The major differences between ModDrop and ModDrop\Plus\Plus{} are zoomed in on the upper-right panels. With multiple available modalities (columns 1-3), we find that there are fewer false positives (FP, green) and false negatives (FN, red) for ModDrop\Plus\Plus{} than for ModDrop. When there is only a single modality available (column 4), more FNs can be observed in both ModDrop and ModDrop\Plus\Plus{}, but ModDrop\Plus\Plus{} still produces better results than ModDrop with fewer FPs and FNs.

\begin{table}[t]
\centering
\caption{Quantitative results on ISBI 2015 dataset. The best and second best performances are denoted by \textbf{bold} and \underline{underline}.}
\rowcolors{2}{gray!25}{white}
\begin{tabular}{|cccc|cccc|cccc|}
\hline
\multicolumn{4}{|c}{Modalities} & \multicolumn{4}{|c}{DSC($\uparrow$)}                                                                              & \multicolumn{4}{|c|}{SC($\uparrow$)}     \\ \hhline{|====|====|====|}
\multicolumn{1}{|c}{T1}    & \multicolumn{1}{c}{Flair}    & \multicolumn{1}{c}{T2}   & \multicolumn{1}{c|}{PD}   & \multicolumn{1}{c}{MD} & \multicolumn{1}{c}{MD\Plus{}} & \multicolumn{1}{c}{MD\Plus\Plus{}} & \multicolumn{1}{c|}{IM} & \multicolumn{1}{c}{MD} & \multicolumn{1}{c}{MD\Plus{}} & \multicolumn{1}{c}{MD\Plus\Plus{}} & \multicolumn{1}{c|}{IM} \\
\hline
\like{10} & \like{10} & \like{10} & \like{10} & 0.733 & 0.756 &  \underline{0.774} & \textbf{0.793} & 0.717 & 0.747 &  \underline{0.771} & \textbf{0.800} \\
\like{0}  & \like{10} & \like{10} & \like{10} & 0.719 & 0.752 &  \underline{0.768} & \textbf{0.789} & 0.707 & 0.737 &  \underline{0.766} & \textbf{0.779} \\
\like{10} & \like{0}  & \like{10} & \like{10} & 0.691 & 0.694 &  \underline{0.695} & \textbf{0.735} & 0.717 & 0.723 &  \underline{0.730} & \textbf{0.750} \\
\like{10} & \like{10} & \like{0}  & \like{10} & 0.725 & 0.743 &  \underline{0.768} & \textbf{0.775} & 0.713 & 0.738 &  \underline{0.760} & \textbf{0.775} \\
\like{10} & \like{10} & \like{10} & \like{0}  & 0.725 & 0.750 &  \underline{0.774} & \textbf{0.778} & 0.701 & 0.742 &  \underline{0.767} & \textbf{0.787} \\
\like{0}  & \like{0}  & \like{10} & \like{10} & 0.684 & 0.693 &  \underline{0.695} & \textbf{0.727} & 0.708 & 0.727 &  \underline{0.734} & \textbf{0.748} \\
\like{0}  & \like{10} & \like{0}  & \like{10} & 0.710 & 0.735 &  \underline{0.766} & \textbf{0.772} & 0.708 & 0.727 &  \underline{0.772} & \textbf{0.776} \\
\like{0}  & \like{10} & \like{10} & \like{0}  & 0.710 & 0.742 &  \underline{0.772} & \textbf{0.782} & 0.693 & 0.726 &  \underline{0.772} & \textbf{0.773} \\
\like{10} & \like{0}  & \like{0}  & \like{10} & 0.668 &  \underline{0.679} & 0.677 & \textbf{0.720} & 0.697 &  \underline{0.733} & 0.729 & \textbf{0.739} \\
\like{10} & \like{0}  & \like{10} & \like{0}  & 0.684 & 0.688 &  \underline{0.696} & \textbf{0.731} & 0.692 & 0.715 &  \underline{0.727} & \textbf{0.748} \\
\like{10} & \like{10} & \like{0}  & \like{0}  & 0.717 & 0.747 &  \underline{0.777} & \textbf{0.779} & 0.704 & \underline{0.715} &  \textbf{0.793} & \textbf{0.793} \\
\like{10} & \like{0}  & \like{0}  & \like{0}  &  \underline{0.656} & 0.633 & 0.616 & \textbf{0.717} & 0.675 &  \underline{0.700} & 0.685 & \textbf{0.749} \\
\like{0}  & \like{10} & \like{0}  & \like{0}  & 0.680 & 0.709 &  \underline{0.742} & \textbf{0.773} & 0.684 & 0.715 &  \underline{0.770} & \textbf{0.782} \\
\like{0}  & \like{0}  & \like{10} & \like{0}  & 0.680 & 0.686 &  \underline{0.694} & \textbf{0.715} & 0.680 & 0.712 &  \underline{0.721} & \textbf{0.736} \\
\like{0}  & \like{0}  & \like{0}  & \like{10} & 0.647 & 0.649 &  \underline{0.655} & \textbf{0.682} & 0.674 &  \underline{0.724} & 0.716 & \textbf{0.725} \\
\hline

\end{tabular}
\end{table}

\section{Discussion and Conclusion}
The main advantages of ModDrop\Plus\Plus{} lie in the following two aspects: (1) it allows us to train one unified model for all missing conditions, which is applicable and efficient in clinical practice. Moreover, a unified model that can take an arbitrary combination of modalities is also a pre-requisite to study transfer learning and domain generalization problems for multi-modal data: otherwise only common modalities from source and target domains can be considered as in \cite{ackaouy2020unsupervised,aslani2020scanner,DLB}, rather than leveraging the rich information from the distinct modalities. (2) with the plug-and-play dynamic head and the intra-subject co-training strategy, ModDrop\Plus\Plus{} can be easily applied to any existing CNN models. 

Other methods for missing modalities exist, such as HeMIS \cite{havaei2016hemis}. However, these methods typically require changing model architectures and the impact on performances caused by such changes may not be trivial, especially when a specific model architecture has been proved very effective on certain applications. Therefore, the focus of our study is to investigate the feasibility of improving a widely used training strategy for missing modality problems (ModDrop) without modifications on model architecture, and minimizing the performance gap between the unified model and the modality-specific models. Though our experiments demonstrate the effectiveness of ModDrop\Plus\Plus{}, we can still observe performance gaps between ModDrop\Plus\Plus{} and modality-specific models from the results on the ISBI dataset, suggesting the potential for further improvement. One limitation of our study is the dataset size, especially the ISBI dataset, where, even though each subject has 4-5 time points, only 5 unique subjects are available. Validation in a larger cohort thus remains as future work.

\section{Acknowledgements}
This work was supported in part by the NIH grant R01-NS094456, in part by the NIH and NIBIB grant T32EB021937, and by the Advanced Computing Center for Research and Education (ACCRE) of Vanderbilt University.

\bibliographystyle{splncs04}

\bibliography{references.bib}

\end{document}